\documentclass[aps,twocolumn,groupedaddress,showpacs]{revtex4}
\usepackage{graphics}
\usepackage{color}
\usepackage{ifthen}
\usepackage{amssymb}
\usepackage{tabularx}

\def\({\left(}
\def\){\right)}
\def\[{\left[}
\def\]{\right]}
\def\be{\begin{equation}}
\def\ee{\end{equation}}
\def\beq{\begin{eqnarray}}
\def\eeq{\end{eqnarray}}
\def\nn{\nonumber}
\def\fnl{f_{\rm NL}}

\def\nn{{\nonumber}}
\newcommand{\dd}{\mathrm d}

\newcommand{\an}{\rm an }

\begin{document}

\title{On Minkowski Functionals of CMB polarization}

\author{Pravabati Chingangbam$^{1,4}$} 
\email{prava@iiap.res.in} 

\author{Vidhya Ganesan$^{1,2}$} 
\email{vidhya@iiap.res.in} 

\author{K. P. Yogendran$^{3}$\footnote{on leave from IISER Mohali, Sector 81, Mohali, India}} 
\email{yogendran@iisertirupati.ac.in} 

\author{Changbom Park$^{4}$} 
\email{cbp@kias.re.kr} 
\affiliation{$^1$ Indian Institute of Astrophysics, Koramangala II Block,
  Bangalore  560 034, India\\
$^2$ Department of Physics, Indian Institute of Science, Bangalore  560 012, India\\
$^3$ IISER Tirupati, Karakambadi Road, Tirupati 517 507, India\\
$^4$Korea Institute for Advanced Study, 85 Hoegiro, Dongdaemun-gu,
Seoul 02455, Korea} 

\begin{abstract}
CMB polarization data is usually analyzed using $E$ and $B$ modes because they are scalars quantities under rotations along the lines of sight and have distinct physical origins.  We explore the possibility of using the Stokes parameters $Q$ and $U$ for complementary analysis and consistency checks in the context of searches for non-Gaussianity. We show that the Minkowski Functionals (MFs) of $Q,U$ are invariant under local rotations along the lines of sight even though $Q,U$ are spin-2 variables, for full sky analysis. The invariance does not hold for incomplete sky. 
For local type  primordial non-Gaussianity, when we compare the non-Gaussian deviations of MFs for $Q,U$ to what is obtained for $E$ mode or temperature fluctuations, we find that the amplitude is about an order of magnitude lower and the shapes of the deviations are different. This finding can be useful in distinguishing local type non-Gaussianity from other origins of non-Gaussianity in the observed data. 
Lastly, we analyze the sensitivity of the amplitudes of the MFs for $Q$, $U$ and the number density of singularities of the total polarization intensity to the tensor-to-scalar ratio, $r$, and find that all of them decrease as $r$ increases. 

\end{abstract}


\maketitle
\section{Introduction} 

Towards the last stages of the epoch of recombination quadrupolar anisotropies must have been present in the intensity of photons. This anisotropy would lead to a net linear polarization of the Cosmic Microwave Background (CMB)  photons~\cite{Rees:1968,Bond:1987ub,Crittenden:1993wm,Frewin:1994,Coulson:1994,Crittenden:1995} as a result of Thompson scattering with free electrons. 
In addition to the temperature fluctuations, the CMB polarization is a vital repository of clues about the physical properties, origin of primordial fluctuations and history of the Universe. 
In the standard inflationary~\cite{Starobinsky:1979,Guth:1981,Starobinsky:1982,Linde:1982,Albrecht:1982} $\Lambda$CDM cosmology the quadrupolar anisotropies can be traced back to two physical origins, namely, anisotropies in the scalar density fluctuations of the plasma, and tensor perturbations of the metric. 

Observations of polarized CMB photons measure the Stokes parameters $Q$ and $U$ along each line of sight. They transform as spin-2 objects under rotations about the line of sight. Using spin two spherical harmonics they can be re-expressed as the so-called $E$ and $B$ modes~\cite{Zaldarriaga:1996xe,Kamionkowski:1996ks}. To first order in perturbations, $E$ mode is sourced by scalar density perturbations while $B$ mode is sourced by tensor perturbations. Due to polarization taking place only during the last stages of decoupling the rms of $E$ mode is about one order of magnitude lower than that of temperature fluctuations. $E$ mode has been measured~\cite{Kovac:2002fg} and used for cosmological  analysis~\cite{Kogut:2003et,Ade:2015xua,Ade:2015ava}. 
Generic inflationary models predict that the ratio of the amplitudes of the primordial tensor and scalar perturbations, denoted by $r$, is less than one, with the precise value being model dependent. Currently, the detection of $B$ mode sourced by primordial tensor perturbations is one of the foremost goals of observational cosmology. The precise knowledge of its rms value, which translates into knowledge of $r$, will strongly constrain inflation models. From observations by BICEP2 and KECK Array the latest constraint on $r$ is $<0.07$ at 95\% CL~\cite{Array:2015xqh}. 


Inflation predicts that the fluctuations in the  energy density and metric during the very early stages of the Universe are random variables with a nearly Gaussian probability distribution function. The statistical properties of these fluctuations are inherited by the temperature fluctuations and polarization of the CMB.  One of the important tools to analyze the statistical properties of these random fields are the Minkowski
Functionals (MFs)~\cite{Tomita:1986,Coles:1988,Gott:1990,Mecke:1994,Schmalzing:1997, Schmalzing:1997uc,Winitzki:1997jj,Matsubara:2003yt}. They are 
quantities that characterize the geometrical and topological properties of excursion sets of the CMB fields. They have been applied to temperature fluctuations to constrain primordial non-Gaussianity~\cite{Komatsu:2011,Planck:2013} and more 
recently on the $E$ mode polarization~\cite{Vidhya:2014,Ade:2015ava}. They have also been used to identify traces of residual foreground contamination in WMAP data~\cite{Chingangbam:2012wp}. 


In this paper we focus attention on MFs for CMB polarization. 
Most analysis of polarization data focus on using $E$ and $B$ modes since they are scalar quantities under rotations along the lines of sight and the clean separation of their physical origins. Their invariance under rotations means that their MFs are invariant under such rotations. 
When dealing with observed data, $E,B$ are obtained from the directly observed $Q,U$ variables using  spin-2 spherical harmonic functions. This step is complicated by the fact that we need to work with incomplete sky coverage of the data. 
Given that MFs are real space quantities it is then a natural question to ask whether it is possible to extract the physical information that we seek from MFs of $E,B$ equally well from $Q,U$.

An early work that uses the genus, which is one of the MFs, for $Q$ and $U$ can be found in~\cite{Changbom_Changyung}.  
Because of their spin-2 nature it is not immediately obvious whether their MFs measured by different observers (or experiments) can be meaningfully compared. We clarify this issue and show analytically that for full sky coverage their  MFs are invariant under rotations along the line of sight. We then confirm the invariance by performing numerical calculations of MFs. The invariance breaks down for incomplete sky. 

Next, we investigate how non-Gaussian deviations of primordial density fluctuations manifest in MFs of $Q,U$. We restrict our analysis to local type primordial non-Gaussianity. We find, in comparison to what is obtained for $E$ mode and temperature fluctuations, the non-Gaussian deviations of the MFs for $Q,U$ have corresponding amplitudes that are about an order of magnitude lower, but the deviation shapes are distinct. 
Analytic expressions for MFs and the number density of singularities of the total polarization intensity, $P$, for Gaussian primordial perturbation, are derived in~\cite{Naselsky:1998by, Dolgov:1999pa}. The amplitudes of the MFs and the number density of singularities are expressed in terms of the variances of $Q,U$. Their invariance under rotations along the lines of sight relies on the above result. We analyze the effect of tensor perturbations on the amplitudes of the MFs and the number density of singularities and find that both decrease as the amplitude of tensor perturbations is increased. This result is useful for searches for $B$ mode and consistency checks of the CMB data.  
We would like to mention that we do not take into account observational effects, such as beam effect and instrument noise, in this paper since our goal is to elucidate the theoretical issues. It is expected that when such realistic effects are included the statistical significance of the results on the non-Gaussian deviations and the sensitivity to the presence of primordial tensor perturbations will weaken.

This paper is organized as follows. In section II we briefly describe the CMB polarization simulation that we use for our analysis. In section III we analyze the effect of rotations of the coordinate axes along the line of sight on the variances of $Q$ and $U$. 
In section IV we introduce MFs, discuss their numerical computation and demonstrate their invariance under global rotations of the coordinate axes along the lines of sights. Further, we calculate the MFs for $Q$ and $U$ containing input primordial non-Gaussianity. 
In section V we present the effect of including $B$ mode on the variances and the amplitude of MFs of $Q$ and $U$. We also show the effect on the number of singularities of the total polarization intensity. 
We end by summarizing the results 
along with a discussion of their implications in section VI.

\section{CMB polarization fields and their simulations}

The Stokes parameters $Q$ and $U$ transform as spin-two objects under a rotation by angle $\alpha$ about each line of sight, given by
\begin{equation}  
\left(
\begin{array}{c}
Q'\\ U'
\end{array}
\right) = R(2\alpha)\left(
\begin{array}{c}
Q\\ U
\end{array}
\right) =
\left( 
\begin{array}{c}
\cos 2\alpha \ Q + \sin 2\alpha\ U\\
-\sin 2\alpha \ Q + \cos 2\alpha\ U
\end{array}
\right).
\label{eqn:qutransf}
\end{equation}
Equivalently, Eq.~(\ref{eqn:qutransf}) can be written as $\left( Q'\pm iU'\right) =e^{\mp i2\alpha}\left(Q\pm iU\right)$. 
They are related to the total polarization intensity as, $P\equiv \sqrt{Q^2+U^2}$ and the polarization angle as, $\varphi \equiv \frac{1}{2}\tan^{-1}U/Q$. $P$ is invariant under rotations about the line of sight while $\varphi$ is not.
Further, $Q$ and $U$ can be expressed as $E$ and $B$ modes~\cite{Zaldarriaga:1996xe} by expanding  $Q\pm iU$ in terms of spin-two spherical harmonics, $Y_{\pm 2,\ell m}$, 
\begin{equation}
Q\pm iU = \sum_{\ell m} a_{\pm 2,\ell m}Y_{\pm 2,\ell m},
\label{eqn:qu2eb}
\end{equation}
and defining
\begin{eqnarray}
a_{E,\ell m} &=& -\frac12 \left(a_{2,\ell m}+a_{-2,\ell m}\right) \nn\\
a_{B,\ell m} &=& \frac{i}{2} \left(a_{2,\ell m}-a_{-2,\ell m}\right),
\label{eqn:aeb}
\end{eqnarray}
and 
\begin{eqnarray}
E(\hat n) &=& \sum \bigg(\frac{\ell+2}{\ell-2}\bigg)^{1/2} a_{E,\ell m} Y_{\ell m} \nn\\
B(\hat n) &=& \sum\bigg(\frac{\ell+2}{\ell-2}\bigg)^{1/2} a_{B,\ell m} Y_{\ell m}. 
\end{eqnarray}
It is useful for our subsequent analysis to invert Eq.~(\ref{eqn:aeb}). Inserting $a_{\pm 2,\ell m}$ into Eq.~(\ref{eqn:qu2eb}) gives
\begin{eqnarray}
Q &=& -\frac12 \sum_{\ell m} \bigg\{ a_{E,\ell m}\left(Y_{2,\ell m} + Y_{-2,\ell m}\right)  \nn \\
{} && \hskip 1.2cm + i a_{B,\ell m}\left(Y_{2,\ell m} - Y_{-2,\ell m}\right) \bigg\} \nn \\
U &=& \frac{i}2 \sum_{\ell m} \bigg\{ a_{E,\ell m}\left(Y_{2,\ell m} - Y_{-2,\ell m}\right)  \nn \\
{}&& \hskip 1.2cm + i a_{B,\ell m}\left(Y_{2,\ell m} + Y_{-2,\ell m}\right) \bigg\}.
\label{eqn:queb}
\end{eqnarray}

For our analysis we produce simulations of $E$ and $B$ mode with Gaussian statistics for primordial scalar and tensor perturbations and corresponding input angular power spectra. The $\Lambda$CDM cosmological parameters values used are $\Omega_ch^2=0.1198,\ \Omega_bh^2=0.02225,\ H_0=67.27,\ n_s=0.9645,\  ln(10^{10}A_s)=3.094,\ \tau=0.079$, taken from the 2015 PLANCK data~\cite{Ade:2015xua},  The input angular power spectra were obtained using the \texttt{CAMB} package~\cite{Lewis:2000ah,cambsite} and the map simulations were made using the \texttt{HEALPIX} package~\cite{Gorski:2005,Healpix}. The map resolution corresponds to HEALPIX parameter NSIDE value 1024. The amplitude of $B$ mode maps are fixed by choosing values of the tensor-to-scalar ratio, $r$. Maps of $Q$, $U$ are then made using the $E$ and $B$ maps. 
\section{Variances of $Q,U$ and $\nabla Q,\nabla U$}

In this section we examine the transformation of the variances of $Q,U$ and their gradients $\nabla Q, \nabla U$, under rotations along the line of sight.  
Let us denote the the following variances of field $X$ by
\begin{equation}
\Sigma_0^{X} \equiv \langle XX\rangle, \quad \Sigma_1^{X} \equiv \langle \nabla X\cdot\nabla X\rangle ,
\label{eqn:sigma}
\end{equation}
where $X$ can be either $Q$ or $U$. Note that $\langle\rangle$ here means averaging over the surface of the sphere. 

For simplicity, we first consider rotation by the same angle $\alpha$ along every line of sight. Then from Eq.~(\ref{eqn:qutransf}) we get
\begin{eqnarray}
\langle Q'Q'\rangle &=& \cos^2(2\alpha)\, \langle QQ\rangle + \sin^2(2\alpha)\, \langle UU\rangle \nn\\
{} && +\sin 4\alpha\ \langle QU\rangle, \label{eqn:qq} \\
\langle U'U'\rangle &=&  
\sin^2(2\alpha)\, \langle QQ\rangle + \cos^2(2\alpha)\, \langle UU\rangle \nn \\
{}&& - \sin 4\alpha\ \langle QU\rangle, \label{eqn:uu}\\
\langle \nabla Q'\cdot\nabla Q'\rangle &=&  \cos^2(2\alpha)\,\langle \nabla Q\cdot \nabla Q\rangle 
+\sin^2(2\alpha)\,\langle \nabla U\cdot \nabla U\rangle    \nonumber  \label{eqn:dqdq}\\
{} && + \sin 4\alpha\ \langle \nabla Q\cdot\nabla U\rangle,\\
\langle \nabla U'\cdot\nabla U'\rangle &=&  \sin^2(2\alpha)\,\langle \nabla Q\cdot \nabla Q\rangle +\cos^2(2\alpha)\,\langle \nabla U\cdot \nabla U\rangle \nonumber\\
&& - \sin 4\alpha\ \langle \nabla Q\cdot\nabla U\rangle,
\label{eqn:dudu}
\end{eqnarray}
Eqs.~(\ref{eqn:qq})-(\ref{eqn:dudu}) imply that $\langle QQ\rangle \ne \langle Q'Q'\rangle $ and $\langle \nabla Q\cdot \nabla Q\rangle \ne \langle \nabla Q'\cdot \nabla Q'\rangle$ if the cross-correlations $\langle QU\rangle$ and $\langle \nabla Q\cdot\nabla U\rangle$ are non-zero. 
Using Eq.~(\ref{eqn:queb}) we get 
\begin{eqnarray}
\langle QU\rangle &=& \frac{i}4 \sum_{\ell m \ell' m'} \bigg\{  a_{E,\ell m} a^{\ast}_{E,\ell' m'} \int d\Omega \left(Y_{2,\ell m} + Y_{-2,\ell m}\right) \nn\\
{}&& \hskip 1.2cm  \times \left(Y^{\ast}_{2,\ell' m'} - Y^{\ast}_{-2,\ell' m'}\right)  
 \nonumber\\
&& 
+ a_{B,\ell m} a^{\ast}_{B,\ell' m'}  
\int d\Omega\left(Y_{2,\ell m} - Y_{-2,\ell m}\right) \nn \\
{}&& \hskip 1.2cm \times \left(Y^{\ast}_{2,\ell' m'} + Y^{\ast}_{-2,\ell' m'}\right)  \nonumber\\ {} && + {\rm cross \ terms}\big\} 
\end{eqnarray}

Let $\langle\rangle_{\rm ens}$ denote averaging over an ensemble of Universes.  
Using the following relations which follow from isotropy of the linear perturbations,
\begin{eqnarray}
\big\langle a_{E,\ell m} a^{\ast}_{E,\ell' m'} \big\rangle_{\rm ens} &=& \delta_{\ell \ell'}\delta_{m m'} \big\langle |a_{E,\ell m}|^2 \big\rangle_{\rm ens} 
\label{eqn:aee_ensemble}\\
\big\langle a_{B,\ell m} a^{\ast}_{B,\ell' m'} \big\rangle_{\rm ens} &=& \delta_{\ell \ell'}\delta_{m m'} \big\langle |a_{B,\ell m}|^2 \big\rangle_{\rm ens} 
\label{eqn:abb_ensemble}\\
\big\langle a_{E,\ell m} a^{\ast}_{B,\ell' m'} \big\rangle_{\rm ens} &=& 0, 
\label{eqn:aeb_ensemble} 
\end{eqnarray}
the ensemble average of $\langle QU\rangle$ becomes 
\begin{eqnarray}
\big\langle\langle QU\rangle\big\rangle_{\rm ens} &=& \frac{i}4 \sum_{\ell m} \big\{  \big\langle |a_{E,\ell m}|^2 \big\rangle_{\rm ens}  \int d\Omega \nn \\
{}&& \left(Y_{2,\ell m} + Y_{-2,\ell m}\right) \left(Y^{\ast}_{2,\ell m} - Y^{\ast}_{-2,\ell m}\right)  
 \nonumber\\
&& 
+ \big\langle |a_{B,\ell m}|^2 \big\rangle_{\rm ens}   \int d\Omega \nn \\
{} &&\left(Y_{2,\ell m} - Y_{-2,\ell m}\right)  \left(Y^{\ast}_{2,\ell m} + Y^{\ast}_{-2,\ell m}\right) \big\}.
\label{eqn:qu_ensemble2}
\end{eqnarray}
$Y_{s,\ell m}$ satisfy the conjugacy relation
\begin{equation}
Y_{s,\ell m}^{\ast} = (-1)^{m+s} Y_{-s, \ell -m},
\label{eqn:Yconjugacy}
\end{equation}
where $s$ is the spin index. 
Using the conjugacy relation  and the reality condition for $a^{\ast}_{E,\ell m}$, we get
\begin{eqnarray}
 \big\langle\langle QU\rangle\big \rangle_{\rm ens} &=&  \frac{i}4 \sum_{\ell m} \big\{  \big\langle |a_{E,\ell m}|^2 \big\rangle_{\rm ens} 
-  \big\langle |a_{B,\ell m}|^2 \big\rangle_{\rm ens} \big\} \nonumber\\
{} && \times\int d\Omega \big(-Y_{2,\ell m} Y^{\ast}_{-2,\ell m} \nn \\
{} && \ \ + \,Y_{-2,\ell m}Y^{\ast}_{2,\ell m}\big).
\label{eqn:qu_ens} 
\end{eqnarray}
The two terms in the integrand above are complex conjugates. Again using the conjugacy relation we get 
\begin{equation}
Y_{2,\ell m} Y^{\ast}_{-2,\ell m} = (-1)^{m-2}\,Y_{2,\ell m} Y_{2,\ell -m} 
\end{equation}
Since the dependence of $Y_{2, \ell m}$ on the coordinates $\theta, \phi$ is like $Y_{2,\ell m} = f(\theta)e^{im\phi}$ and  $Y_{2,\ell -m}=g(\theta) e^{-im\phi}$, each term in the integrand of Eq.~(\ref{eqn:qu_ens}) is real. Therefore,  the relative sign in the integrand leads to the two terms cancelling. Thus we get 
\begin{equation}
\big\langle\langle QU\rangle\big\rangle_{\rm ens}=0.
\label{eqn:qucor}
\end{equation}
Since the cancellation occurs before we have integrated over the sky, we actually have
\begin{equation}
\big\langle QU\big\rangle_{\rm ens}=0,
\end{equation}
which holds at every point $(\theta,\phi)$.
Using Eq.~(\ref{eqn:qucor}) in Eq.~(\ref{eqn:qq}) we get 
\begin{equation}
\big\langle\Sigma_0^{Q'}\big\rangle_{\rm ens}
 = \cos^2(2\alpha)\, \big\langle\langle QQ\rangle\big\rangle_{\rm ens} + \sin^2(2\alpha)\, \big\langle\langle UU\rangle \big\rangle_{\rm ens}.
\end{equation}
Using $\langle QQ\rangle_{\rm ens} = \langle UU\rangle_{\rm ens}$ we get 
\begin{equation}
\langle\Sigma_0^{Q'}\rangle_{\rm ens}=\langle\Sigma_0^{Q}\rangle_{\rm ens}. 
\end{equation}
And similarly for $U$.

Next, to calculate $\langle \nabla Q\cdot\nabla U\rangle$ we need to simplify the factor containing gradients of the $Y's$, given below,
\begin{eqnarray}
\nabla\left(Y_{2,\ell m} + Y_{-2,\ell m}\right)\cdot\nabla \left(Y^{\ast}_{2,\ell m} - Y^{\ast}_{-2,\ell m}\right)  
\nn \\
= \nabla Y_{2,\ell m}\cdot\nabla Y^{\ast}_{2,\ell m} - \nabla Y_{-2,\ell m}\cdot\nabla Y^{\ast}_{-2,\ell m} \nonumber \\
 -\nabla Y_{2,\ell m}\cdot\nabla Y^{\ast}_{-2,\ell m} + \nabla Y_{-2,\ell m}\cdot\nabla Y^{\ast}_{2,\ell m} \nonumber. \\
&&
\end{eqnarray}
The first two terms cancel using the conjugacy relation in Eq.~(\ref{eqn:Yconjugacy}). So we are left with
\begin{eqnarray}
 \big\langle\langle \nabla Q\cdot\nabla U\rangle\big\rangle_{\rm ens} &=&  \frac{i}4 \sum_{\ell m} \bigg\{  \big\langle |a_{E,\ell m}|^2 \big\rangle_{\rm ens} 
\nn\\
{}&& -  \big\langle |a_{B,\ell m}|^2 \big\rangle_{\rm ens} \bigg\} 
\int d\Omega \nn \\
&& \bigg(-\nabla Y_{2,\ell m}\cdot \nabla Y^{\ast}_{-2,\ell m} \nn \\
&& + \nabla Y_{-2,\ell m}\cdot \nabla Y^{\ast}_{2,\ell m}\bigg)  
\end{eqnarray}
Again using the conjugacy relation, and using the fact that the dependence of $Y_{2,\ell m}$ on $\phi$ is $e^{im\phi}$ while that of $Y_{2,\ell -m}$ is $e^{-im\phi}$, we can show that each term inside the integrand is a real function and hence the two terms cancel. 
Therefore,
\begin{equation}
\big\langle\langle \nabla Q\cdot\nabla U\rangle\big\rangle_{\rm ens}=0.
\label{eqn:dqducor}
\end{equation}
Again, the zero correlation holds at every $(\theta, \phi)$. 

We have shown that the variances are invariant under a {\em global} rotation by the same angle about every line of sight. If we allow the rotation angle to vary for different lines of sight and retrace the above calculation, then the rotation factors will be part of the integrand over the sphere. However, in order to prove Eqs. (\ref{eqn:qucor}) and  (\ref{eqn:dqducor}) we only use the properties of $Y_{\pm 2,\ell m}$ and do not need to carry out the integration at any step. Therefore, the invariance holds for direction dependent rotations also.

In the case of incomplete sky due to Galactic and point sources masks, the relations  (\ref{eqn:aee_ensemble}) and (\ref{eqn:abb_ensemble}) no longer hold because isotropy is broken. Therefore, in this case
$$
\big\langle\langle QU\rangle\big\rangle_{\rm ens}\ne 0, \quad 
\big\langle\langle \nabla Q\cdot\nabla U\rangle\big\rangle_{\rm ens}\ne 0,$$
which implies that $\Sigma_0^{X}$ and $\Sigma_1^{X}$ are not invariant under rotations along the lines of sight.  

\section{Minkowski Functionals} 

A useful way to study the statistical properties of a random field is to choose suitable threshold values of the field and analyze regions that have field values above each threshold. Such regions are called {\em excursion sets}. 
The morphological properties of the excursion sets  and their variation with the threshold value can reveal the statistical nature of the field.
 The shapes of the excursion sets can be quantified 
in terms of geometrical and topological quantities, namely, the
Minkowski Functionals (MFs).
There are three MFs
for two-dimensional manifolds such as the excursion sets of the
CMB. The first, denoted by $V_0$, is the area fraction of the
excursion set. The second, denoted by $V_1$, is the total length of
iso-temperature contours or boundaries of the excursion set. The
third, denoted by $V_2$, is the genus which is the difference between
the numbers of hot spots and cold spots. Let $f$ denote a generic random field, $\sigma_0=\sqrt{\Sigma_0}$ is the rms of $f$, and let $u\equiv f/\sigma_0$. Let $\nu$ be the threshold value chosen from the range of $u$. Then, the MFs are defined  mathematically as follows:
\begin{equation}
V_0(\nu)\equiv \int \dd a,\quad V_1(\nu)\equiv\frac14\int_C \dd l,\quad V_2(\nu) \equiv \frac{1}{2\pi}\int_{C} K \,\dd l,
\label{eqn:mfformula}
\end{equation}
where $\dd a$ is the area element of the excursion set, $C$ denotes contours that form the boundaries of the excursion sets, $\dd l$ is the line element on $C$ and $K$ is the curvature of the contours. 
Closely related to the MFs are the two Betti numbers~\cite{Park:2013dga, Chingangbam:2012km}, of which the first is the number of the connected regions and the second is the number of holes in the connected regions. The difference of the first and second Betti numbers gives the genus. 

In this section, we study the effect of rotations along the line of sight on the MFs of $Q$ and $U$. Further, we analyze how primordial non-Gaussianity will show up in the MFs of $Q$ and $U$. For all calculations shown in this section we set the $B$ mode to be zero.

\subsection{Numerical computation of Minkowski Functionals for random fields}

In general, for a given random field we may not know the analytic form for the MFs or we may need to test whether an observed random field has the statistical property that we expect from theory. In such situations we need to calculate MFs using numerical methods. For the numerical calculation of MFs we employ the method due to Schmalzing and Gorski~\cite{Schmalzing:1997uc}, which relies on expressing Eq.~(\ref{eqn:mfformula}) as a discrete sum involving first and second order covariant derivatives on the sphere and a $\delta$-functional of the field. The discretization of the $\delta$-function introduces numerical inaccuracies~\cite{Lim:2012}. For a given random field, let $V_i$ and $V^{\rm an}_i$ denote the numerically calculated MFs and the exact value, respectively. Then we can write $V_i=V^{\rm an}_i + R_i^{\Delta \nu}$, where $R_i^{\Delta \nu}$ denotes the residual numerical error. The superscript $\Delta\nu$ is the threshold bin size and it indicates that the residual error is dependent on it. 
$R_i^{\Delta \nu}$ is given by
\begin{equation}
R_i^{\Delta\nu}(\nu)=\frac{1}{\Delta\nu}\int_{
\nu-\Delta\nu/2}^{\nu+\Delta\nu/2} \dd u 
V_i(u) \ -\ V^{\an}_i(\nu).
\label{eqn:mferror}
\end{equation}
 
\subsection{Minkowski functionals for Gaussian $Q$ and $U$}

\begin{figure}
\begin{center}
\resizebox{2.3in}{2.3in}{\includegraphics{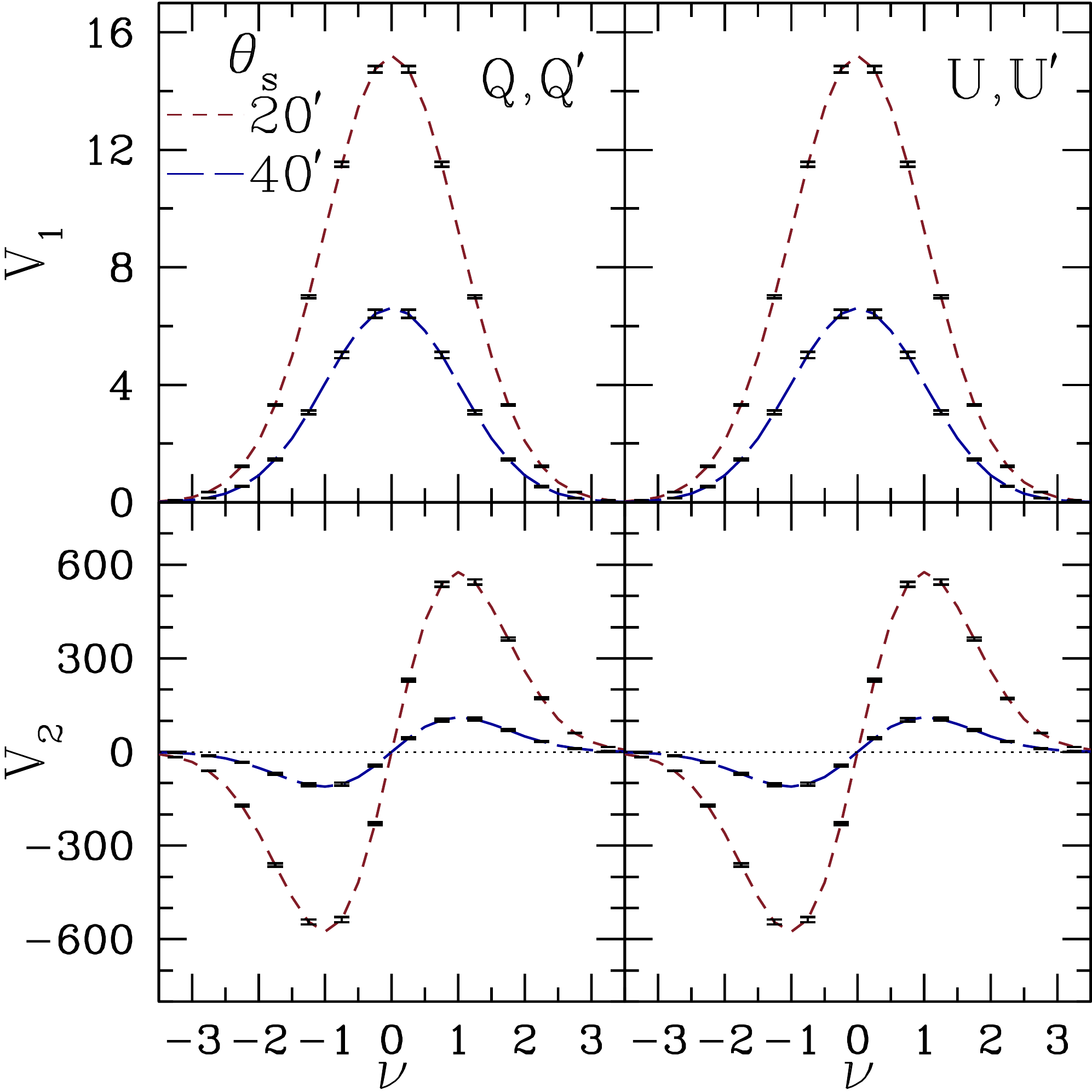}}
\end{center}
\caption{Upper panels show plots of the contour length, $V_1$, for Gaussian $Q,Q'$ (left) and $U,U'$ (right), for smoothing angles $\theta_s=20'$ (red) and $\theta_s=40'$ (blue). $Q', U'$ have been obtained by rotating $Q,U$ about each line of sight by angle $\alpha=45^{\circ}$. The two  plots in each panel are indistinguishable, demonstrating numerically that the amplitudes are invariant under global rotations along the line of sight. All plots are average over 1000 simulations and the error bars are the sample variances. Lower panels show the genus, $V_2$. We have repeated the calculations for other rotation angles and the results remain the same.}
\label{fig:qu_gmf}
\end{figure}

\begin{figure*}
\begin{center}
\resizebox{3.3in}{3.3in}{\includegraphics{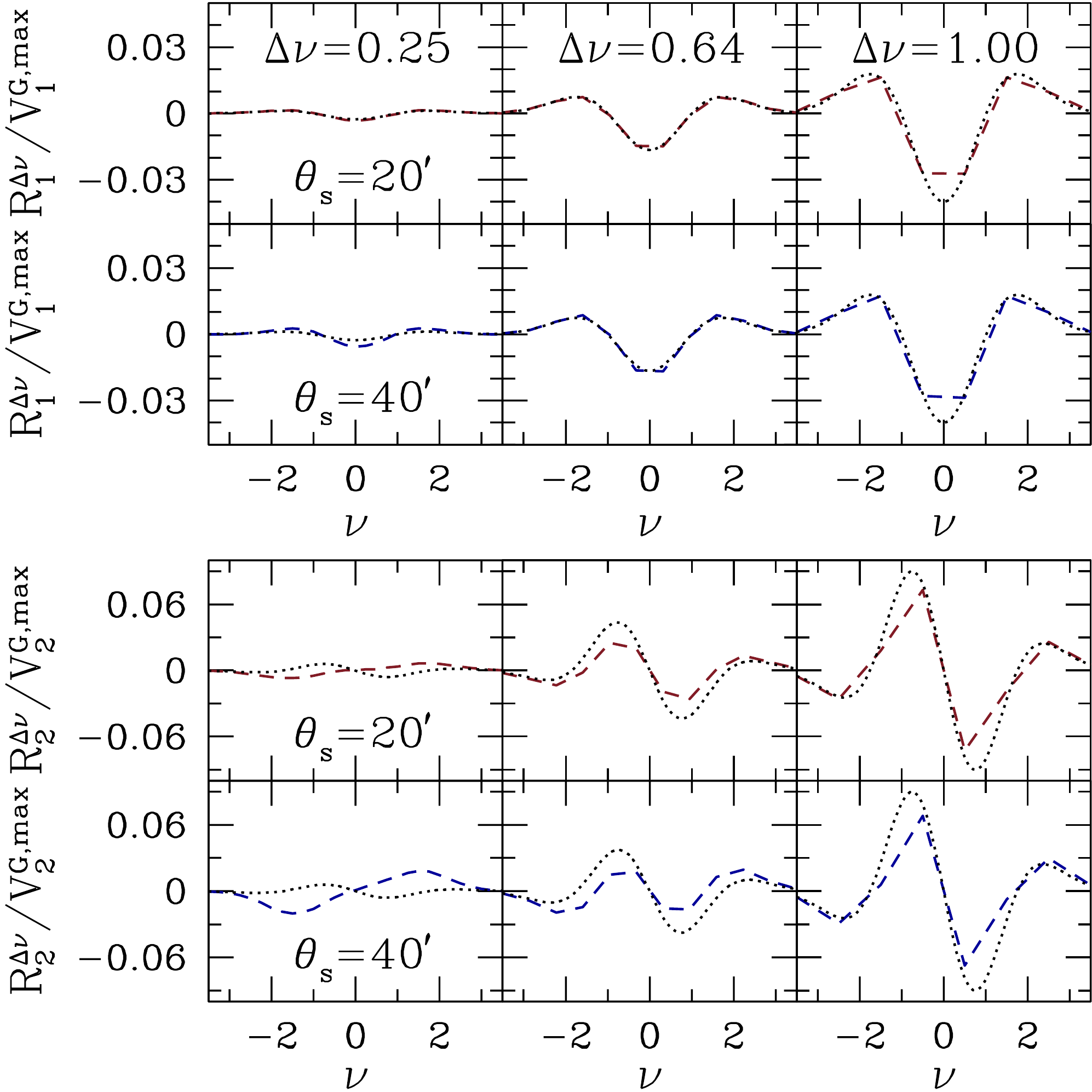}}
\resizebox{3.3in}{3.3in}{\includegraphics{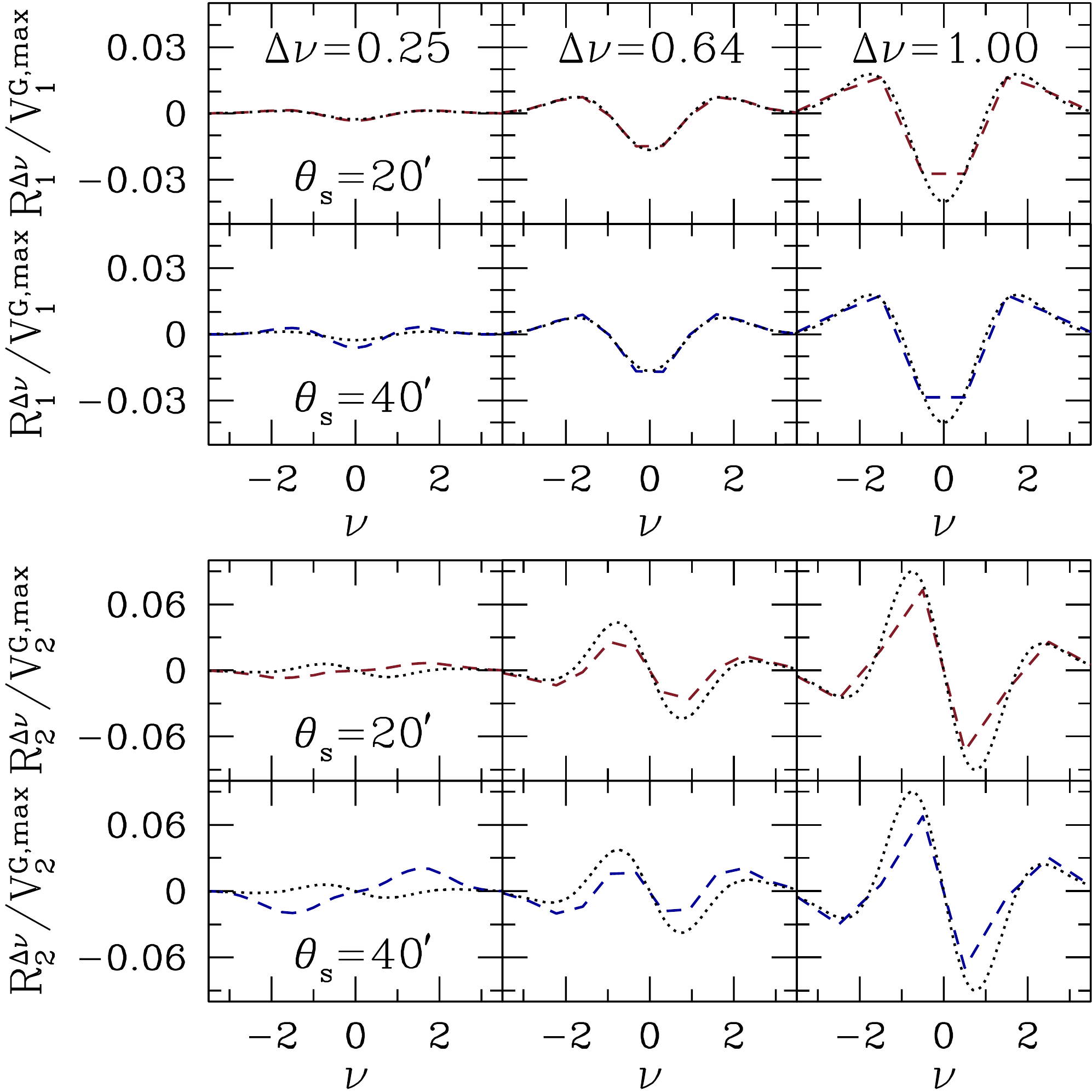}}
\caption{{\em Left}: Upper panels show of the residuals defined in Eq.~(\ref{eqn:mferror}) of $V_1$, obtained numerically for Gaussian $Q$ for smoothing angles $\theta_s=20'$ (red) and $\theta_s=40'$ (blue) for $\Delta\nu=0.25,0.64, 1$. 
All plots are average over 1000 maps. The black dotted lines are the analytic form of the residuals obtained by integrating the first term on the right hand side of Eq.~(\ref{eqn:mferror}) for a Gaussian {\em scalar} field. 
  Lower panels show the residuals for $V_2$. {\em Right}: Same as left plots for $U$. }
\label{fig:qu_residual}
\end{center}
\end{figure*}
For a Gaussian random field, $f$, the MFs as functions of the threshold $\nu$, are given by
\begin{equation}
\label{eqn:gmf}
V_k(\nu)=A_k \, H_{k-1}(\nu)\,e^{-\nu^2/2}, \quad k=0,1,2.
\end{equation}
Here $H_k(\nu)$ is the $k$-th Hermite polynomial and the amplitudes are 
\begin{equation}
\label{eqn:ak}
A_0=\frac{1}{\sqrt{2\pi}},\quad A_1=\frac1{8}
\frac{\sigma_1}{\sqrt{2}\sigma_0},\quad A_2=\frac1{(2\pi)^{3/2}}
\left(\frac{\sigma_1}{\sqrt{2}\sigma_0}\right)^2,
\end{equation}
where $\sigma_1=\sqrt{\Sigma_1}$ is the rms of the gradient of the field.


If the primordial scalar and tensor fluctuations are Gaussian, we expect $Q$ and $U$ to be Gaussian for linear perturbations and their MFs should be of the form given in Eq.~(\ref{eqn:gmf}). The ratio $r_c\equiv\sigma_0/\sigma_1$ 
 is usually referred to as the correlation length of hot and cold structures in a given random field. 
The amplitudes of the MFs depend on powers of $r_c$. As shown in section III, $\sigma_i^X$, and hence $r_c$, are invariant under  rotations by the same angle about every line of sight. Hence the MFs should be invariant. However, for incomplete sky they are not invariant. Our subsequent analyses focus on full sky calculations.  

We have computed the MFs for simulated Gaussian $Q,U$ and their corresponding $Q',U'$ obtained by rotating $Q,U$ by angle $\alpha$. In Fig.~(\ref{fig:qu_gmf}) we plots of the contour length and the genus for Gaussian $Q,Q'$ and $U,U'$ for rotation angle $\alpha=45^{\circ}$, demonstrating their invariance. We show results for two smoothing angles $\theta_s=20', 40'$. We have obtained the same result for other choices of $\alpha$. $B$ mode has been set to zero for these calculation.

The residual error defined in Eq.~(\ref{eqn:gmf}) calculated numerically for $Q$ and $U$ (dashed lines) and the analytic form obtained from integrating the first term of Eq.~(\ref{eqn:gmf}) for a Gaussian {\em scalar} field (black dotted lines), are shown in Fig.~(\ref{fig:qu_residual}). We chose two smoothing angles $\theta_s=20', 40'$. The bin sizes used are $\Delta\nu=0.25, 0.65, 1$. The analytic form of the residuals grow larger for larger bin sizes and are not affected much by changes in the smoothing scale. 
It is interesting to note that the residual errors for the contour length for $Q$ and $U$ are nearly the same and seems to agree with what is expected from Eq.~(\ref{eqn:mferror}) for Gaussian scalar fields (see Fig.~(2) or (3) of~\cite{Lim:2012}). If we zoom in the figure we find that at small bin sizes there is noticeable difference between the dashed and dotted lines. The difference gets more pronounced at larger smoothing angles. They agree very well at larger bin size, such as can be seen for the case $\Delta\nu=1$. 
The genus residuals exhibit similar behaviour but we find much stronger disagreement between the dashed lines and dotted lines at small bin sizes.  Moreover, there is small but noticeable difference between the residuals of $Q$ and $U$. 

We would like to mention that we have repeated the residual calculations for temperature maps and we have reproduced the results of~\cite{Lim:2012}) very well. Hence we rule out the possibility that the disagreement at small bin size arises due to mistakes in our numerical calculations. We think that this disagreement is due to the spin-two nature of $Q$ and $U$.  The derivatives that we have used to implement the Schmalzing and Gorski method are covariant derivatives for scalar fields on the sphere, and not what should be the  appropriate covariant derivative for a $U(1)$ bundle on the sphere which should be relevant for spin-two variables such as $Q,U$.  We do not pursue this question further since the mathematics that is relevant for clarifying it is beyond the scope of this paper. 


\begin{figure*}
\begin{center}
\resizebox{5.5in}{5.in}{\includegraphics{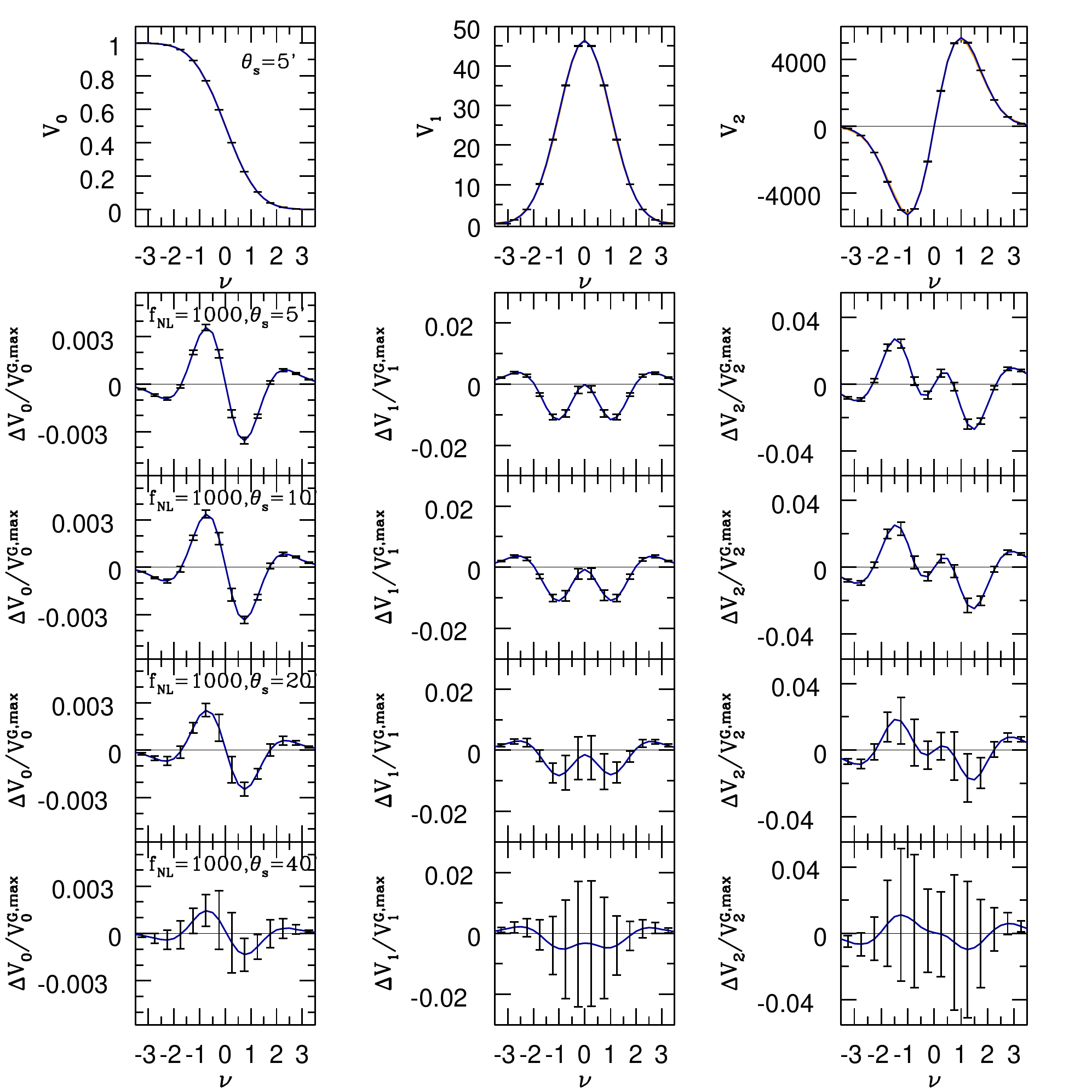}}
\end{center}
\caption{{\em Top panels}: The three Minkowski Functionals for Gaussian and  non-Gaussian cases with $\fnl=1000$ are shown. The plots are not distinguishable by eye since the differences are small. 
{\em Lower panels}: Non-Gaussian deviations for the MFs for $\fnl=1000$ for different smoothing angles. We have chosen unrealistically large value of $\fnl$ because the plots become noisy for small values of $\fnl$. 
 All plots are average over 1000 simulations and the error bars are their sample variances.}
\label{fig:qu_ngmf}
\end{figure*}

Next, let the non-Gaussian deviations of the MFs be denoted by
\begin{equation}
\Delta V_i \equiv V_i^{NG} - V_i^{G}, 
\label{eqn:dmf}
\end{equation}
where $i=0,1,2$. 
In the following we consider local type primordial non-Gaussianity parametrized by the variable $\fnl$. Our calculations here are done for full sky since our aim is to bring out the non-Gaussian effects. Considering incomplete sky will decrease the statistical significance of our results. We use simulations of $a_{E,\ell m}$ that contain input local type primordial non-Gaussianity that have been  made publicly available by Elsner and Wandelt~\cite{Elsner:2009md}. The values of the input $\Lambda$CDM parameters are those obtained from the WMAP 5 years data~\cite{WMAP5}. The resolution is set by NSIDE=512. 
Gaussian and non-Gaussian $Q,U$ maps with our chosen $\fnl$ values are constructed from the corresponding $a_{E,\ell m}$. These maps are then used to calculate the MFs from which $\Delta V_i$ are calculated. We have done the calculations using both the Schmalzing and Gorski method and a geometrical method described in~\cite{Gay:2011wz, Ducout:2012it} and the results agree with each other. 

In Fig.~(\ref{fig:qu_ngmf}) we show the MFs for $Q$ and their non-Gaussian deviations. We have not shown the results for $U$ since they are the same, as expected. The top panels show $V_i$ for Gaussian (black) as well as non-Gaussian maps for $\fnl=1000$, for smoothing angle $\theta_s=5'$. The plots are indistinguishable by eye. The lower panels show the non-Gaussian deviations rescaled by the corresponding $V_i^{G,\rm max}$. 
From the plots of $\Delta V_i$ we can make two main observations. Firstly, the amplitude of deviations for $Q$ is much smaller than what is obtained for temperature fluctuations (compare  Fig.~(\ref{fig:qu_ngmf}) with Fig~(2) of~\cite{Hikage:2006fe}) and for $E$ mode  (compare with Fig~(5) of~\cite{Vidhya:2014}). We have chosen unphysically large values of $\fnl$ because the numerical calculation for realistic values become quite noisy. Secondly,  the shape of deviations is different from what is seen for temperature fluctuations and $E$ mode. The genus deviation is similar to what is seen for cubic order local primordial non-Gaussianity (see Fig.~(4) of~\cite{Chingangbam:2009vi} and Fig.~(1) of~\cite{Matsubara:2010te}.

\section{Effect of primordial tensor perturbations on Minkowski Functionals}

The probability distribution function (PDF) for the total polarization intensity, $P$, has the Rayleigh form $\frac{1}{\sigma_0}P\,e^{-P^2/2\sigma_0}$, where $\sigma_0$ is the rms of $Q$ or $U$, under the assumption that they are equal. As shown in Section III, this holds only for complete sky coverage.  
For the subsequent discussions we consider only complete sky coverage and will refer to $\sigma_0, \sigma_1$ without the field superscript. 
In~\cite{Vidhya:2014} the authors have shown that the PDF of $P$ varies significantly with the amplitude of $B$ mode. 
 Since the PDF of $P$ is completely characterized by $\sigma_0$, for Gaussian $Q,U$, we can quantify the effect of including $B$ mode by calculating how it affects $\sigma_0$. Here we take this observation further and study the effect  on $\sigma_0, \sigma_1$, the amplitude of the MFs for $Q,U$, and the number density of singularities of $P$. 
We use the tensor-to-scalar ratio, $r$, to quantify the effect of primordial tensor perturbations, and study how the various quantities vary as $r$ is varied. 

\begin{figure*}
\begin{center}
\resizebox{2.in}{2.3in}{\includegraphics{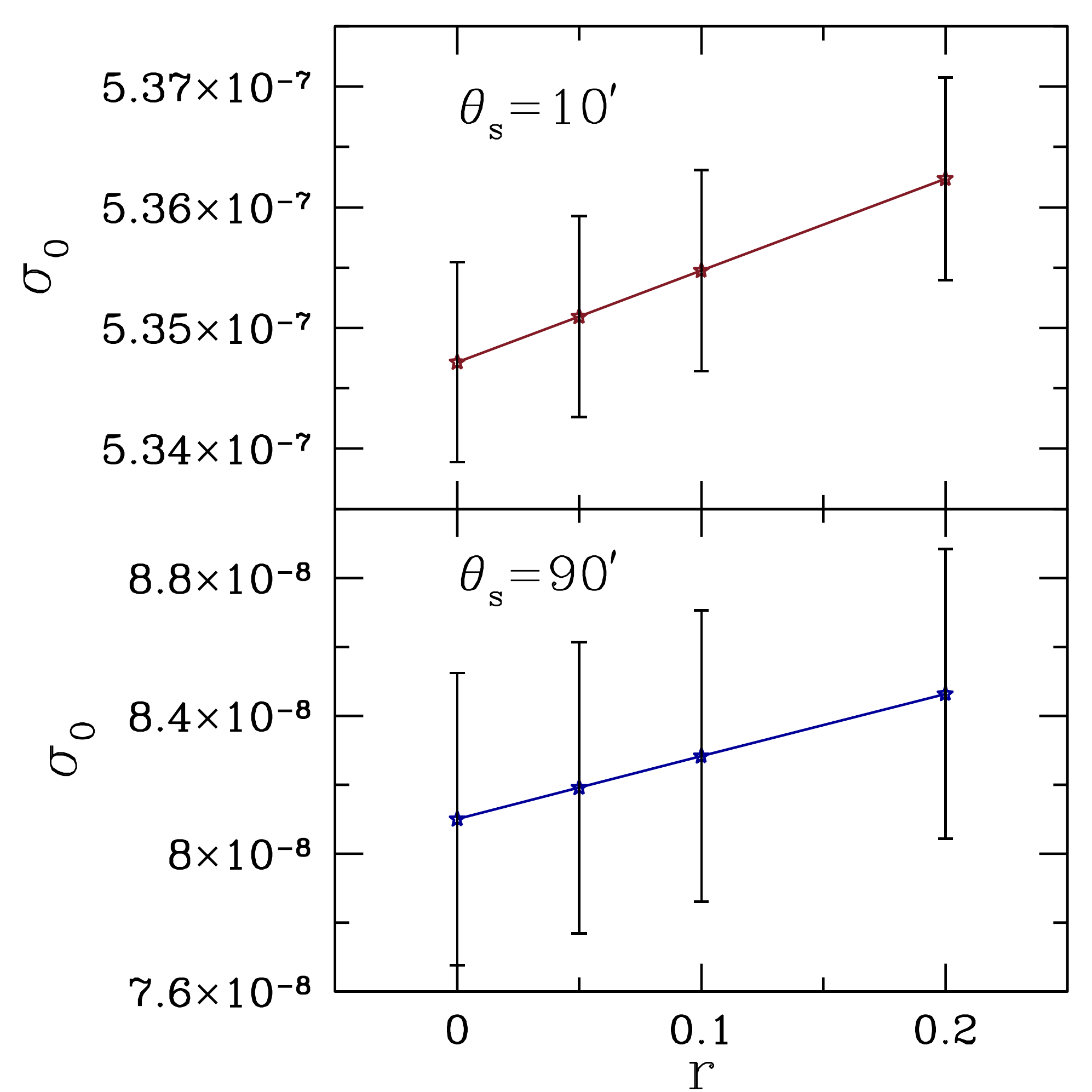}}\quad
\resizebox{2.in}{2.3in}{\includegraphics{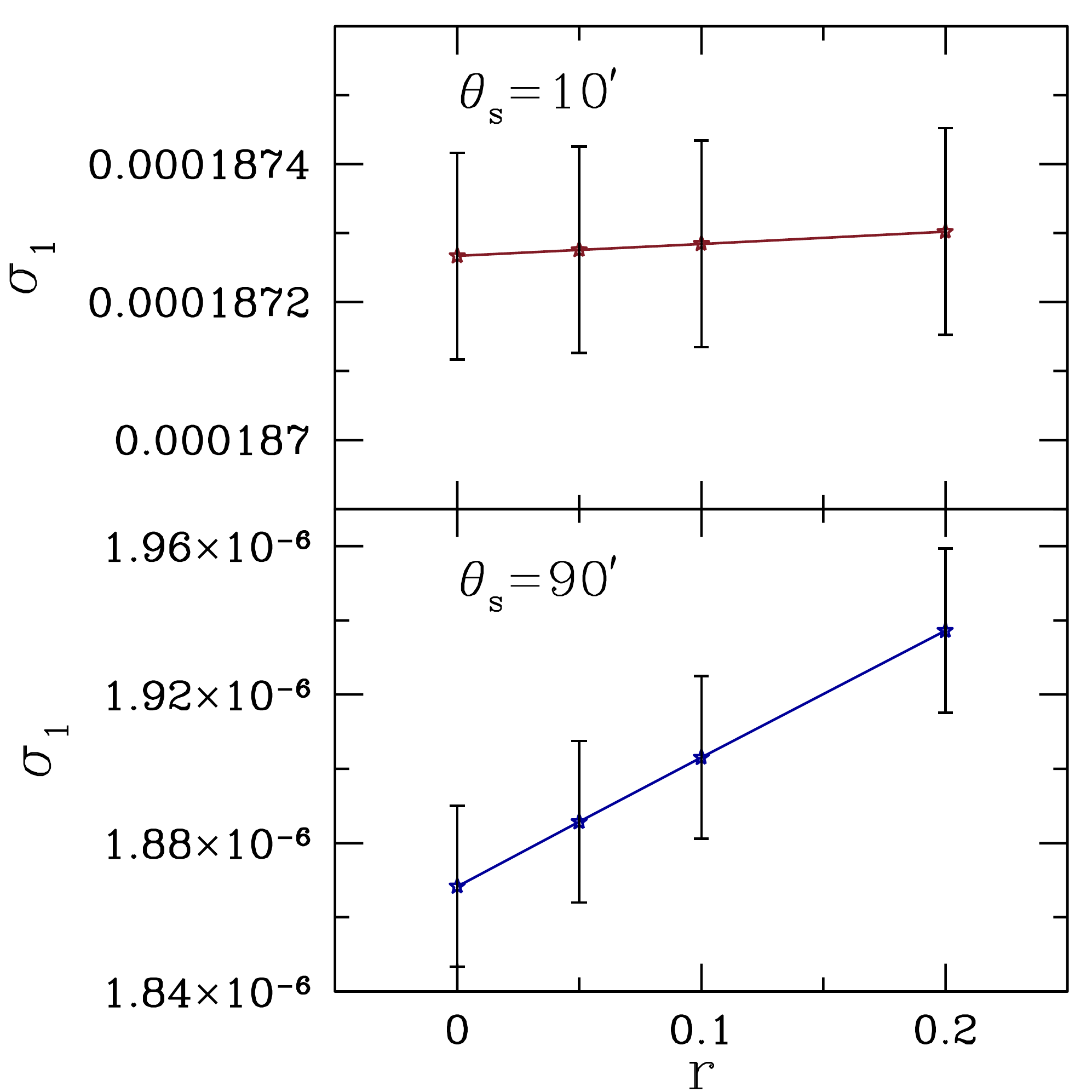}}
\resizebox{2.in}{2.3in}{\includegraphics{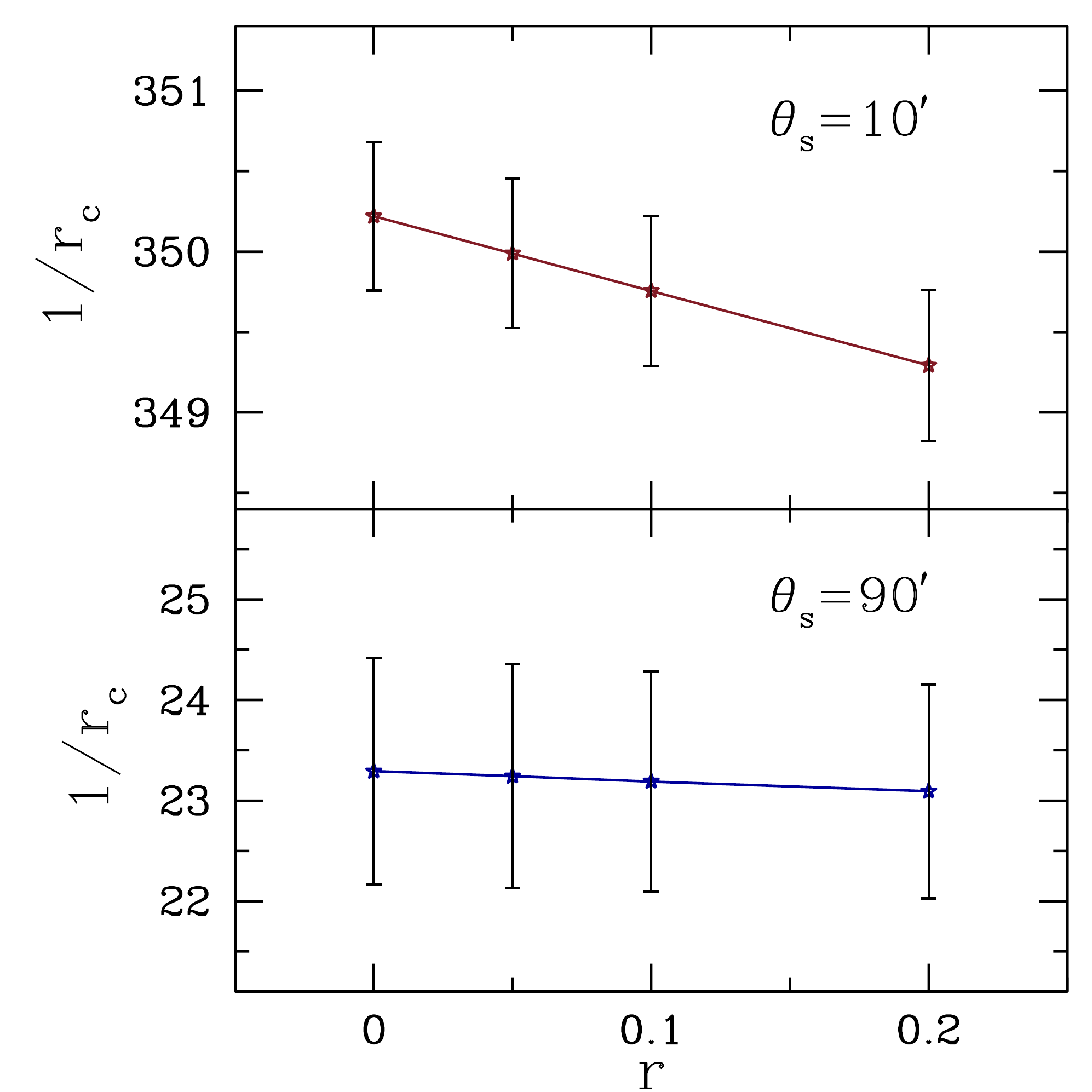}}\quad
\end{center}
\caption{ The left panel shows how $\sigma_0$ varies as $r$ is varied, for smoothing angles $\theta_s=10',90'$.  The middle and right panels show  $\sigma_1$ and $r_c$. The stars indicate values of $r$ at which we have done the calculations. The superscript $Q$ has been dropped. All plots are average over 1000 simulations and the error bars are the sample variances.}
\label{fig:s01_rc_vs_r}
\end{figure*}

 In the left and middle panels of Fig.~(\ref{fig:s01_rc_vs_r}) we show how $\sigma_0$ and $\sigma_1$, computed using Eqn.~(\ref{eqn:sigma}) on simulated maps, vary with $r$ between $0.05$ to 0.2. We have used two smoothing scales, 10' and 90' to highlight the variation in the variances with the smoothing scale also. We  find that inclusion of $B$ mode increases the variances. Note that the slopes of the dependence on $r$ vary with the smoothing scale. 
The dependence of $\sigma_0$ and $\sigma_1$ on $r$ are not linear, even though over the small range that we have considered here they appear to be so on visual inspection. 


To study the effect of $r$ on the amplitudes of the MFs of $Q,U$ it suffices to find out how $r_c$ varies with $r$.  In the right panel of Fig.~(\ref{fig:s01_rc_vs_r}) we have plotted $r_c^{-1}$ versus $r$ between $0.05$ to 0.2, for different smoothing scales. As seen in the plot, the presence of $B$ mode with increasing amplitudes results in decrease of the amplitude of MFs. Note that the genus is more sensitive to $r$ than the contour length. 

For Gaussian $P$, it was shown by Naselsky and Novikov~\cite{Naselsky:1998by} that the amplitudes of the MFs are proportional to $r_c^{-i}$, where $i=1,2$. The behaviour of $r_c^{-1}$ noted above implies that the amplitudes of MFs for $P$ also decrease as we increase $r$.

The points where $Q=0=U$ and hence $P=0$, are referred to as singular points. Let us denote the number density of singularities by $N_{\rm sing}$. In ~\cite{Naselsky:1998by}, it is shown to be given by $N_{\rm sing}=1/4\pi r_c^2$. 
Therefore, our calculation here shows that $N_{\rm sing}$ is sensitive to $r$ and decreases as $r$ is increased. 
This corroborates the result in~\cite{Huterer:2004wt} where the authors found that $N_{\rm sing}$ is sensitive to changes in $r$. Here we have quantified the nature of the dependence.

\section{Conclusion}

Analyses of CMB polarization data in the form of the Stokes parameters $Q,U$ can provide information that complement the analyses using $E,B$ and serve as consistency checks of the results. $Q,U$ are what are measured in CMB polarization observations and hence if we use them for cosmological analysis we bypass the complications related to incomplete sky coverage that arise when we transform to $E,B$ variables.

In this paper we address issues related to analyzing $Q,U$ in the context of searches for primordial non-Gaussianity using MFs. It makes sense to use $Q,U$ for cosmological analysis only if the observable quantities that we define using them are invariant under the transformations under which they transform  as  spin-two variables.  We first  show analytically that under rotations along the lines of sight the MFs are invariant. 
 This implies that calculations of MFs for $Q,U$ can be meaningfully compared between different observers (or different observing instruments) and the physical information obtained from them should be the same. 
 However, we find that the invariance holds only when there is full sky coverage and under the assumption of statistical isotropy of the fluctuations. Thus the result is not immediately applicable to data from actual experiments where parts of the sky are masked due to our uncertain knowledge of Galactic emissions and point sources.

 We have further calculated non-Gaussian deviations of the MFs that arise from local type primordial non-Gaussianity. We find the magnitudes of the deviations are about an order of magnitude lower in comparison to what is seen for $E$ mode or temperature fluctuations and the shapes are distinct.
 For non-Gaussian analysis using masked observed data  MFs an important step is to estimate the Gaussian component. This is usually done by using the Gaussian formulae given in Eq. (\ref{eqn:gmf}), with the amplitudes calculated from the variances of the field obtained from the data. This
 is approximate and is reasonable only for very weakly non-Gaussian fields. The non-Gaussian deviation can then be calculated by subtracting the Gaussian estimate from the MFs calculated from the data.
 Note that this can be applied to MFs of $Q,U$ for masked data. However, for the estimation of error bars, calculations using observed data and simulation can be compared only if the simulations use as input the $x-y$ coordinate choices along each line of sight that is used by the observational setup. 
 

Lastly, we analyze the effect of the presence of primordial tensor perturbations on MFs for $Q$, $U$ and $P$. We also discuss the effect on the number density of singularities in $P$. All these quantities can be expressed in terms of $r_c=\sigma_0/\sigma_1$. We show that $r_c$ is sensitive to the presence of primordial tensor perturbations, and increases as $r$ is increased.
This result can potentially be used in analyzing polarization data and the search for $B$ mode.

\section*{Acknowledgment}
The computation required for this work was carried out on the Hydra
cluster at the Indian Institute of Astrophysics. We acknowledge use of the \texttt{CAMB}~\cite{Lewis:2000ah,cambsite} and {\texttt{HEALPIX}}
packages~\cite{Gorski:2005, Healpix} which were used to derive some of
the results. We thank Anne Ducout and Dmitri Pogosyan for providing us the \texttt{CND\_REG2D}  source code for computing the Minkowski Functionals.  We also thank F.~Elsner for the use of his simulations of non-Gaussian $E$ mode polarization.

\end{document}